# Role of the host genetic variability in the influenza-A virus susceptibility – a review.

**Ana Carolina Arcanjo[1,*], Giovanni Mazzocco[2,3,*], Silviene Fabiana de Oliveira[1,4], Dariusz Plewczynski[2,4,5,6], and Jan P. Radomski[2,7,&]**

*PREPRINT (accepted for the Special Flu Issue in the Acta Biochimica Polonica)*

[1] *Laboratório de Genética, Departamento de Genética e Morfologia, Instituto de Ciências Biológicas, Universidade de Brasília, 70910-900, Brasília, DF, Brazil.* [2] *Interdisciplinary Center for Mathematical and Computational Modeling, Warsaw University, Pawińskiego 5A, Bldg. D, PL–02106 Warsaw, Poland.* [3] *Institute of Computer Science, Polish Academy of Sciences, Warsaw, Poland.* [4] *The Jackson Laboratory for Genomic Medicine, c/o University of Connecticut Health Center; 263 Farmington Avenue, Farmington, CT 06030, USA.* [5] *Yale University, New Haven, CT, USA.* [6] *Center for Innovative Research, Medical University of Bialystok, Poland.* [7] *Institute of Biotechnology and Antibiotics, Starościńska 5, PL–02516 Warsaw, Poland.*

*both authors contributed equally,  &e-mail: janr@icm.edu.pl;*

## Abbreviation List

ADCC – antibody-dependent cell-mediated cytotoxicity
AM – alveolar macrophage
APC – antigen-presenting cell
ASC – apoptosis-associated speck-like protein
ASL – airway surface liquid
C1QBP – component 1 binding protein
CARD – caspase activation and recruitment domain
CD32 – cell surface receptor
CD55 – gene, decay-accelerating factor 55
CMAS – gene, cystidine monophosphate N-acetylneuraminic acid synthase
CMP – cystidine monophosphate
CRD – carbohydrate recognizing domain
CTD – C-terminal domain
CTL – cytotoxic T lymphocyte
DC – dendritic cell
DDX58 – gene, encodes RIG-I
DHX58 – gene, encodes LGP2
DNAJC3 – gene, encodes P58$^{IPK}$
EIF2AK2 – gene, encoded PKR
ER – endoplasmic reticulum
HLA – human leukocyte antigen; human version of the major histocomatibility complex (MHC)
IAV – influenza-A virus
IFIT – interferon-induced proteins with tetratricopeptide
IFN – interferon
IgA – immunoglobulin A
IgG – Immunoglobulin G
IgM – immunoglobulin M
LGP2 – laboratory of genetics and physiology 2 protein
LRR – leucine-rich-repeat
MAVS – mitochondrial antiviral-signaling protein
MUC – gene, mucin
NBD – nucleotide-binding domain
Neu5Ac – sialic acid (also known as neuraminic acid or N-acetylneuraminic acid)
NK – natural killer cell
NLR – NOD-like receptor
NOS2 – nitric oxide synthase 2
NP – influenza virus nucleoprotein
PAMP – pathogen-associated molecular pattern
PKR – ssRNA-binding protein kinase
PRR – pattern recognition receptor
PYD – pyrin domain
RIG-I – retinoic acid-inducible gene I product
RLR – RIG-I-like receptor
RPAIN – replication protein A interacting protein
SLC35A2 – gene, solute carrier family 35 type A2
ST3GAL – gene, sialyltransferase that binds sialic acid to a 2.3 positioned galactose
ST6GAL – gene, sialyltransferase that binds sialic acid to a 2.6 positioned galactose
TAP – transporter of antigen processing
TCR – T-cell receptor
TLR – toll-like receptor
TNF – tumor necrosis factor
TRIM25 – tripartite-motif-containing protein 25
VISA – gene, encodes MAVS

## Abstract

The aftermath of influenza infection is determined by a complex set of host-pathogen interactions, where genomic variability on both viral and host sides influences the final outcome. Although there exists large body of literature describing influenza virus variability, only a very small fraction covers the issue of host variance. The goal of this review is to explore the variability of host genes responsible for host-pathogen interactions, paying particular attention to genes responsible for the presence of sialylated glycans in the host endothelial membrane, mucus, genes used by viral immune escape mechanisms, and genes particularly expressed after vaccination, since they are more likely to have a direct influence on the infection outcome.

## Introduction

Influenza virus is one of the most important cause of infections of the respiratory tract, with 3–5 million clinical infections and 250,000–500,000 fatal cases annually (Dawood *et al.*, 2012; Simonsen *et al.*, 2013; WHO, 2014). Immunity to influenza virus infection has been a research topic for more than 70 years (Andrewes, 1939) due to the relevant impact the illness has on global health care. Although some aspects of the immunological response to influenza-Are well understood, there are still many open research questions in this field.

The influence of viral genetic variability on infection is undoubtedly the most important and heavily investigated topic (Hatta *et al.*, 2001; Wagner *et al.*, 2002; Uipraserktul *et al.*, 2005; Gambaryan *et al.*, 2006; Bateman *et al.*, 2008; Nicholls *et al.*, 2008; Das *et al.*, 2010; Jayamaran *et al.*, 2012; Sriwilaijaroen & Suzuki, 2012; Thakaramaran *et al.*, 2012; Guarnaccia *et al.*, 2013; Koel *et al.*, 2013; Sun *et al.*, 2013; Thakaramaran *et al.*, 2013; El Moussi *et al.*, 2014; Qi *et al.* 2014; Su *et al.*, 2014). The high antigenic drift rate observed in the influenza virus is one of primary reasons why constantly updated seasonal influenza vaccination is recommended. Genomic variability of the influenza virus has been largely investigated over the years, yet the variability of the host remains a sparsely documented topic, despite its crucial impact on the immune response (Wijburg *et al.*, 1997; Matrosovich & Klenk, 2003; Schmitz *et al.*, 2005; Jayasekera *et al.*, 2007; Koyama *et al.*, 2007; Throsby *et al.* 2008; Ichinohe *et al.*, 2009; Sabbah *et al.*, 2009; Kreijtz *et al.*, 2011; Zhou *et al.*, 2012; Henn *et al.*, 2013; Hertz *et al.*, 2013; Lin & Brass, 2013) and the course of infection. The goal of this review is to explore genomic variability in the host genes that mediate host-pathogen interactions. Implemented genomic approach focuses on the major gene variants found in the 1000 Genomes Project data (the 1000 Genomes Project Consortium, 2010; Clarke *et al.*, 2012), and is aimed at understanding how much variability is present across different human populations, and at identifying conservation level of such genes. We attempted to predict whether the genetic variants found in large human populations affect the specificity of virus binding and subsequent effectiveness of the immune response. Genes related to virus entry into the host cell are responsible for the production and assembly of Neu5Ac-α2,6-Galβ1,4-GlcNAc (human), and the Neu5ac-α2,3-Galβ1,4-GlcNAc (avian) receptors in the host membrane. Both the human and avian influenza-A virus receptor proteins, hemagglutinin (HA) and neuraminidase (NA), interact with the above glycans at the beginning of infection. Sialic acid composition and modifications of amino acids near the binding site for HA, and the subsequent site of interaction with NA, influence the success of infection.

## Virus binding to cell

The human influenza virus attaches to the host cell by binding to Neu5Ac-α2,6-Gal-β1,4-GlcNAc (2,6-linked, in short), a sialic acid modification to glycans that is extremely abundant in humans, especially in the epithelial cells of the upper respiratory tract (Skehel & Wiley, 2000). The mucin that protects the cells in the human upper respiratory tract is equally enriched with Neu5Ac-α2,3-Gal (2,3-linked; Cohen *et al.*, 2013), which is recognized by avian strains of influenza virus (Cone, 2009). Both types of sialylglycans are present in different proportions at different regions of the human respiratory tract, and are essential for a success of the infection by influenza-A (Shynia *et al.*, 2006; Nicholls *et al.*, 2007).

Sialyltransferases are responsible for adding Neu5Ac moiety to the glycans (Cohen & Varki, 2010). Twenty genes coding for sialyltransferase were described in the human genome (Harduin-Lepers *et al.*, 2005). Both α2,3 sialyltransferases (ST3GAL1 to 6) and α2,6 sialyltransferases (ST6GAL1 and 2) are responsible for the sialylation of glycans in the human cell membrane. Some of these genes are expressed only in specific tissues, and different sialyltransferases bind the Neu5Ac to different substrates. For example, ST6GAL1 adds the sialic acid moiety in the α2,6 position to the type-II glycans (Galβ1,4GlcNAc) that are *N*-linked, and sometimes to *O*-linked glycoproteins as well. In contrast, ST6GAL2, which is expressed only in brain and fetal tissues, adds the sialic acid in the α2,6 position to oligossaccharides, but not to proteins (Glaser *et al.*, 2007). Therefore, we selected ST6GAL1, ST3GAL1, ST3GAL2, ST3GAL4 and ST3GAL5 genes as relevant for this review, since those are expressed in the human respiratory tract (Kitagawa & Paulson, 1994; Harduin-Lepers *et al.*, 2001; Glaser *et al.*, 2007).

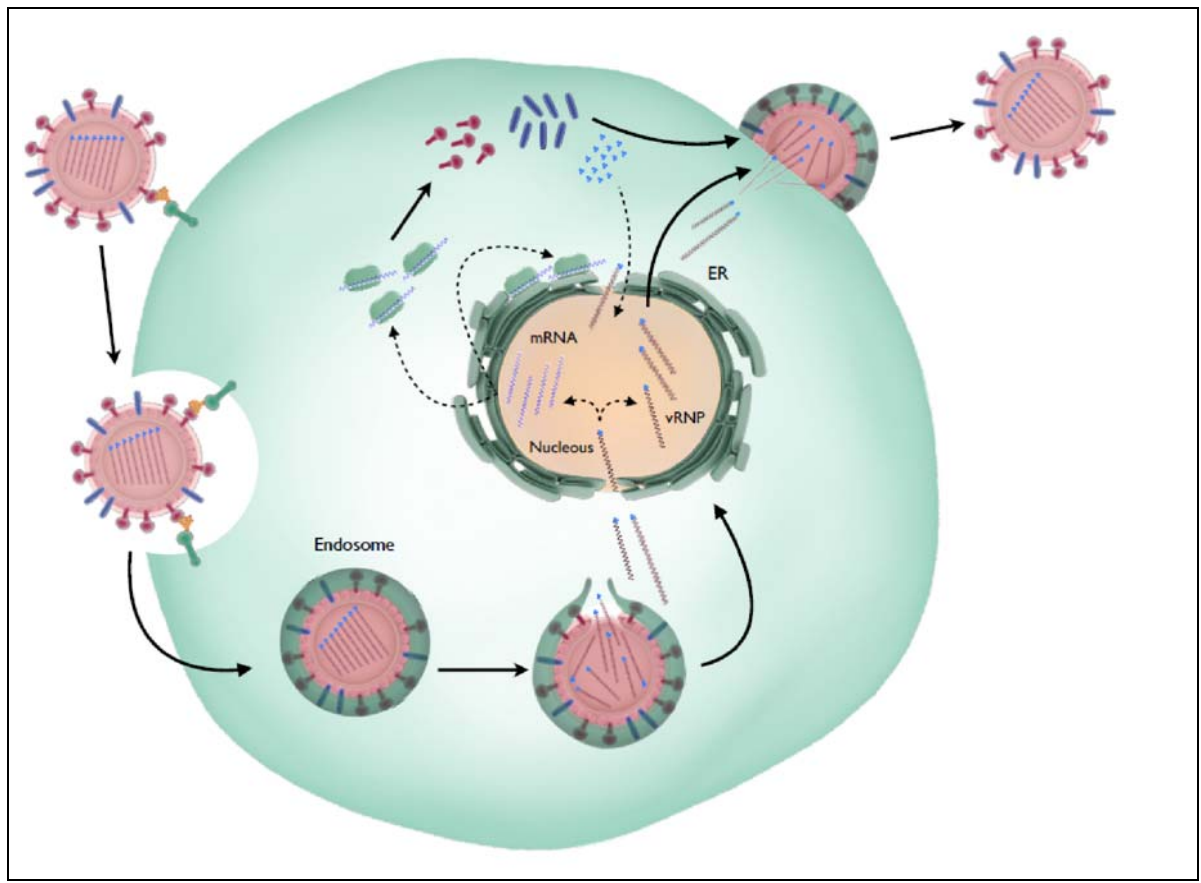

**Figure 1  Influenza virus life cycle**

During the first step of influenza virus infection, the viral glycoprotein HA binds to the receptors on the host cell surface. The virus is then internalized via endocytosis. The low endosomal pH triggers a conformational change in the HA, leading to the fusion of the virus with endosomes and the consequent release of viral ribonucleoprotein complexes (vRNPs) in the cytoplasm. The vRNPs are transported into the nucleus, where both mRNA transcription and vRNA replication/new vRNPs assembly occurs. Translation of viral mRNAs takes place differently in both cytoplasm and endoplasmic reticulum (ER). Following the post-translational modifications (PTM) of certain viral proteins, the structural components of the viral particle are transported to the plasma membrane along with new vRNPs, where viral assembly and virus budding take place.

In order to analyze mechanisms and mutations that can confer complete cell immunity to influenza viruses, the insertional mutagenesis assays were developed in a human cell line (haploid for all chromosomes except chromosome 8). They showed that three independent insertions in the SLC35A2 gene, and two in the CMAS gene are associated with complete cell immunity to influenza viruses (Carette *et al.* 2009). CMAS encodes the enzyme that binds Neu5Ac to a cystidine monophosphate residue. The sialyltransferases cannot transfer Neu5Ac into an acceptor unless it is bound to a CMP residue, because free Neu5Ac cannot be transported to the Golgi membrane (Fleischer, 1998; Munster *et al.*, 1998). SLC35A2 encodes the solute carrier–family 35 transporter type A2, responsible for transporting CMP-Neu5Ac from the cytoplasm to the Golgi membrane.

The glycan receptors are important in cell signaling and adhesion; therefore, they are important for the cell's life cycle. Considering the fundamental role of genes coding for the production of the glycan receptor in susceptibility to influenza, we decided to verify whether the lack of expression of these genes causes any known human disease. Indeed, a malfunction of the SLC35A2 gene is followed by a severe disease: congenital disorder of glycosylation type IIm (Ng *et al.*, 2013). Defects arising from deficient SLC35A2 expression are severe and include developmental delays, hypotonia, variable ocular anomalies, variable brain malformations associated with an unusual serum transferrin profile, seizures, hypsarrhythmia, poor feeding, microcephaly, recurrent infections, dysmorphic features, shortened limbs, and coagulation defects (Ng *et al.*, 2013). All subjects showed diminished proportions of sialic acid glycoconjugates. The profound phenotype and short life span in these patients suggests that malfunctioning of SLC35A2 and/or CMAS is incompatible with life. Therefore, the influenza virus binds to an essential host receptor.

We examined whether variability of the genes responsible for production and assembly of the sialylglycan are indeed more changeable, by searching for the variants described for those genes in the 1000 Genomes ENSEMBL Browser during the pilot phase of the project (1000 Genomes Consortium, 2010). The genes length ranged from 8.8Kb (SLC35A2) to 148Kb (ST6GAL1), with the latter showing 0.5 variants per Kb, whilst the first showed 7.2 variants per Kb. Most of the variants found in any of the analyzed genes were mutations in intronic regions (29–66.7% of described variants), except for the ST6GAL1 gene, for which 72% of variants were located in the 3'UTR region, with 9% variance (*c.f.* Table 1).

**Table 1.** List of variants for seven genes involved in virus binding

| Type of variant* | CMAS | SLC35A2 | ST3GAL1 | ST3GAL2 | ST3GAL4 | ST3GAL5 | ST6GAL1 | Total |
|---|---|---|---|---|---|---|---|---|
| **Frameshift coding** | 0 | 4 | 0 | 2 | 4 | 0 | 0 | **10** |
| **Non-synonymous coding** | 3 | 14 | 4 | 2 | 7 | 11 | 0 | **41** |
| **Splice Site** | 3 | 0 | 4 | 2 | 8 | 5 | 1 | **23** |
| **Synonymous coding** | 1 | 2 | 10 | 2 | 17 | 0 | 9 | **41** |
| **Intronic** | 20 | 30 | 27 | 24 | 67 | 83 | 7 | **258** |
| **5'UTR** | 1 | 1 | 21 | 17 | 1 | 6 | 2 | **49** |
| **3'UTR** | 2 | 10 | 12 | 32 | 6 | 82 | 54 | **198** |
| **Upstream** | 1 | 0 | 0 | 1 | 3 | 5 | 0 | **10** |
| **Downstream** | 2 | 2 | 10 | 2 | 1 | 4 | 2 | **23** |
| **Total Number of Variants** | **33** | **63** | **88** | **84** | **114** | **196** | **75** | **653** |

The genetic variants described in the 1000 Genomes Project Pilot data are: *Frameshift coding:* structural mutation in coding sequence, resulting in a frame shift of reading (**A, B**); *Non-synonymous coding:* nucleotide substitution in the coding sequence (**A, B**), resulting in an amino acid change in the peptide chain (**A, C**); *Splice Site:* 1-3bp into an exon or 3-8bp into an intron (**A**); *Synonymous coding:* nucleotide substitution in coding sequence (**A, B**), but not resulting in an amino acid change (**C**); *Intronic:* mutation in intron (**A**); *5'UTR:* in 5' untranslated region (**A, B**); *3'UTR:* in 3' untranslated region (**A, B**); *Upstream:* mutation within 5kb upstream of the 5' end of transcript (**A**); *Downstream:* mutation within 5kb downstream of the 3' end of transcript (**A**).

SLC35A2 showed both non-synonymous and frame-shift coding variants, which could potentially alter the phenotype of the resulting protein. Interestingly, SLC35A2 was the only gene examined in which malfunction causes a severe and ultimately lethal congenital disorder and the absence of a functional copy of the gene causes severe developmental defects (Ng *et al.*, 2013). Nevertheless, considering the crucial importance of CMP-Neu5Ac transport from the cytoplasm to Golgi apparatus, some studies have shown that SLC35A3 (which is an UDP-Gal/UDP-GlcNAc transporter) can help the transporting of CMP-Neu5Ac in cases where SLC35A2 is working poorly (Olczak *et al.*, 2013). None of the two insertions in the CMAS gene nor the three insertions in the SLC35A2 gene described by Carette *et al.* (2009) were found in the screening of the 1000 Genomes Project data.

Several articles reported a change in binding affinity of the HA molecule to the host receptor that are related to amino acid changes in the structure of HA, which cause a conformational change in the receptor binding site of the protein (Ohuchi *et al.*, 1997; Matrosovich *et al.*, 2000; Hatta *et al.*, 2001; Gambaryan *et al.*, 2006; Bateman *et al.*, 2008; Tria *et al.*, 2013). The binding of influenza-A avian virus (IAV) HA to the sialylglycan (α2,3 or α2,6 linked) is rather weak ($K_{diss} > 10^{-4}$M). As result, successful entry of a virus into a cell is facilitated when several HAs bind to several sialylglycans (Matrosovich & Klenk, 2003). If any variant of analyzed genes affect the availability or the quantity of sialylglycans on the cell surface, it will also affect the success rate of cell infection. In this context, it is known that the topology and density of glycans in the membrane, as well as time of incubation, are also involved in the successful binding of proteins to glycans (Lewallen *et al.*, 2009), and that the differences in specificity of binding of avian and human influenza virus rely not only on α2,3-linked or α2,6-linked sialic acid binding, but also on differences in the structure of the sialylated glycans. Human-like strains of influenza virus bind to glycans that display an open-umbrella shape, which is formed by α2,6 sialic acids bound to long chains of oligossaccharides. On the other hand, avian strains of influenza virus HA bind to α2,3 or α2,6 sialic acid only if they are bound to short chains of oligosaccharides, in a cone shape (Chandraserakan *et al.*, 2008).

The binding of influenza virus and viral entry into a cell (**Fig. 1**) are two early steps during the virus-host interaction, in which human genetic variability can either predispose or protect from infection. How the virus accesses the cell and the numerous host mechanisms preventing virus infection are also subject to genetic changes. One host gene that protects against influenza infection is CD55, coding for a protective decay-accelerating factor that prevents cell damage caused by complement molecules (Osuka *et al.*, 2007). The CD55 gene is expressed in all types of cells in the two forms: a GPI-membrane anchored form (mCD55), which is particularly well expressed in the plasma membrane of all blood tract cells, and a soluble form (sCD55) found in body fluids (Medof *et al.*, 1987). The CD55 is associated with prevention of lung lesions, being largely expressed in this type of tissue (Osuka *et al.*, 2007). A SNP (rs2564978) on the promoter region of the CD55 gene was found to be associated with severe influenza-A H1N1 pandemic 2009 infection in Chinese and Japanese individuals. The rs2564978T/T genotype was found to be highly associated with severe form of influenza, while the rs2564978C/C was associated with a less severe form (Zhou *et al.*, 2012). When searching the frequency of this SNP in the 1000 Genomes Project database, we found that rs2564978T is more frequent in Chinese populations (54% in Chinese from Beijing and 63% in Southern Han Chinese), while in other populations its frequency ranged from 1.7% (Yoruba from Ibadan) to 39% (Japanese from Tokyo). Despite the higher frequency of the protective alleles in Yoruba, the African population was the one that suffered the most from the spread of the influenza-A H1N1 2009 pandemic, with a relative death rate of 7.8 deaths per 100,000 habitants in the first 12 months of circulation of the virus (Dawood *et al.*, 2012). The second severely affected population was the Southern Asians, ranging from 3.3 deaths per 100,000 habitants (considering only respiratory failure), to 4.4 deaths per 100,000 habitants (considering both respiratory and cardiovascular failures associated with influenza). Europeans and Western Pacific individuals showed the smallest rates of deaths per 100,000 habitants (1.8 and 1.7, respectively), and indeed those populations have the highest frequencies of the protective C allele (70–90%).

### *The protective properties of mucus against Influenza-A Viruses*

Epithelial mucus is a dynamic semipermeable barrier that enables exchange of nutrients, water, gases, odorants, hormones and gametes while being impermeable to most bacteria and many pathogens (Cone, 2009). It is present in the respiratory tract, where it is called Airway Surface Liquid (ASL) and constitutes a primary innate barrier to foreign molecules and pathogens (Lillehoj & Kim, 2002). The ASL consists of a strictly regulated mixture of water, salts and various macromolecules such as mucins, proteoglycans, lipids, and other proteins that confer viscosity and defensive function to ASL. Whenever one of these components is missing, or is present at suboptimal concentrations, there is large probability of disease such as asthma and other respiratory diseases (Lillehoj & Kim, 2002). ASL is rich in α2,3-linked and α2,6-linked moieties, both in secreted and membrane-bound forms (Nicholls *et al.*, 2007).

It was suggested that mucins are the most important components of ASL, because of their high concentrations in the secreted portion of the mucus and high number of genes encoding both membrane-bound and secreted forms of mucins expressed in the human genome. Thirteen mucin-encoding genes were described in the human genome, eight of which are expressed in the respiratory tract – products of the five are secreted, and of the three remaining are membrane-bound (Lillehoj & Kim, 2002). Mucins are high-molecular-weight glycoproteins containing variable numbers of amino acid tandem repeats enriched in Ser, Thr and Pro residues (Voynow *et al.*, 1998; Ogasawara *et al.*, 2007). The presence of these repetitive amino acids is responsible for heavy glycosylation and thus polydispersity in both size and charge of mucin molecules. Glycosylation takes place between the Ser/Thr moieties of the peptide backbone and *N*-acetylgalactosamine of the oligosaccharides, that is characteristic of *O*-linked glycoproteins (Lillehoj & Kim, 2002; Rose & Voynow, 2006). 95% of all secreted mucins are the MUC5 type, which is heavily *O*-glycosylated (Lillehoj & Kim, 2002) and is acceptor of both α2,3-linked and α2,6-linked sialic acids. The sialic acid bound to these mucins is responsible for the lubricant and high viscosity features of the mucus (Nicholls *et al.*, 2012, Varki & Gagneux, 2012). MUC2 is a secreted form of the protein that comprise 2.5% of mucins in the ASL (Kirkham *et al.*, 2002; Rose & Voynow, 2006) containing about 40% sialylated oligosaccharides (Karlsson *et al.*, 1997). MUC2 expression is heavily upregulated during inflammatory processes or infection, mainly due to activation by IL-4 and IL-9 (Lillehoj & Kim, 2002). The ASL contains also other defense-related proteins, such as protease-inhibitors, anti-oxidants, proteases, anti-microbial proteins, IgA and cytokines (Lillehoj & Kim, 2002). The high concentrations of IgA in the mucus are indicative of an early infection, suggesting that antibodies and immunoglobulins are secreted in the mucus to fight the initial phases of infection (Kreijtz *et al.*, 2011).

Some proteases were shown to activate influenza virus, and are possibly responsible for the pneumopathogenicity of the virus (Lillehoj & Kim, 2002). Mucins showing anti-protease activity (Nadziejko & Finkelstein, 1994), seem to improve the host defense system. Nicholls *et al.* (2007) hypothesized that the high content of sialylated proteins in the secreted mucus is a physical barrier to the influenza virus, preventing the virus from accessing target epithelial cells. The physiological role of cellular sialic acid receptors is to promote mucus adherence to the epithelial cell surface that protects epithelial tissues from dehydration, microbial pathogens, and reactive oxygen species produced by infectious bacteria and/or the oxidative burst of leucocytes (Ogasawara *et al.*, 2007). However, sialoglycoproteins mediating the cell adherence and viscoelasticity of mucus, and serve as receptor sites for the binding of exogenous macromolecules, including influenza virus HA.

The mucus line of defense comprises a viscoelastic gel up to 50μm thick, immobilizing bacteria and viruses, that are then cleared by the cilia action of ciliated cells of the respiratory tract. The enhanced defence is probably due also to the soluble sialylated proteins in the mucus, that mimic the natural ligands of influenza on the cell surface, that immobilize virus on the mucus layer (Duez *et al.*, 2008; Cohen *et al.*, 2013). They found that the higher the concentration of sialic acids in the mucus, the less epithelial cells were infected *in vitro*, especially if the mucus samples are treated with the neuraminidase (NA) inhibitor oseltamivir phosphate. The virus could find its way through the mucus layer towards epithelial cells in the absence of oseltamivir, since NA is able to free the virus entangled in mucus by disrupting the binding of the virus HA to sialic acid. The rate of virions clearance depends on the interactions of motile cilia from the trachea with the overlaying mucus. It is not yet clear what determines the rates of mucociliary clearance, which vary within proximal and distal regions of the airways. Certainly, the effeicacy of cilia are primary determinants of the basal mucociliary clearance rate, but the quantity and viscoelastic properties of mucin may also be important variables (Knowles & Boucher, 2012).

Moreover, proteins with CRDs are also likely to interact with HA and NA, since those two viral proteins have *N*-linked oligossaccharides side chains. Yang *et al.* (2011) showed that Galectin-1 can bind influenza virus and inhibit viral infection both *in vitro* and in mice. The authors observed that the galectin-1 expression is upregulated in animals during ongoing infection in a viral-load-dependent manner (the larger the viral load, the more galectin-1 expressed). Also, mice that were infected with a neurovirulent strain of influenza-A virus showed a higher rate of survival when treated with galectin-1 intranasally, and with a lower degree of apoptosis and inflammation in the lungs, showing that galectin-1 is an effective inhibitor of viral reproduction and reduces the symptoms of infection.

## *Inside a host cell: innate immune response to influenza virus infection.*

The innate immune system constitutes the first line of defense against a viral infection (**Fig. 2**). During this phase, epithelial host cells can recognize novel influenza infection via pattern recognition receptors (PRRs). PRRs recognize pathogen-associated molecular patterns (PAMPs) expressed during virus replication (Hale *et al.*, 2010). The principal PAMP for influenza is the viral RNA bearing a 5'-triphospate group (PPP-RNA), which constitutes a molecular signature that distinguishes foreign RNA from host RNA (Abbas *et al.*, 2013). The receptors able to sense the viral RNA are: toll-like receptors (TLRs), RIG-I-like receptors (RLRs), NOD-like receptors (NLRs) (Pang & Iwasaki, 2011), and interferon-induced proteins with tetratricopeptide repeats (IFITs, Abbas *et al.*, 2013).

The major function of these receptors is to initiate expression of pro-inflammatory cytokines and type-I interferons: IFN-α and IFN-β (Heil *et al.*, 2004; Lund *et al.*, 2004), that are able to inhibit protein synthesis in host cells, limiting the replication of viruses. They also stimulate dendritic cells (DCs) to enhance both CD4+ and CD8+ antigen presentation, thus promoting activation of the adaptive immune response (Kreijtz *et al.*, 2011). The innate cellular response to viral infection is mainly carried out by macrophages and DCs. Alveolar macrophages (AM) have somewhat opposing effects after becoming activated in the host's lungs. On the one hand, they phagocytose influenza virus-infected cells, thus limiting viral spread (Tumpey *et al.*, 2005; Kim *et al.*, 2008); and on the other, they produce nitric oxide synthase 2 (NOS2) and tumor necrosis factor alpha (TNF-α), thus contributing to influenza virus-induced pathology (Jayasekera *et al.*, 2007; Lin *et al.*, 2008). These competing effects clearly underscore the delicate equilibrium that exists within the immune system. Another AM role is in regulating the immune response, especially in the development of antigen-specific T-cell immunity (Wijburg *et al.*, 1997). Of note, while AM produce only low levels of pro-inflammatory cytokines (van Riel *et al.*, 2011), blood-derived macrophages infected with influenza viruses produce large amounts of pro-inflammatory cytokines.

### *Receptors I: Toll-like receptors (TLR)*

Toll-like receptors (TLRs) are a highly conserved family of PRRs glycoproteins and play a pivotal role in the innate immune recognition of viruses (**Fig. 2**). TLRs consist of an N-terminal extracellular PAMP-binding domain with multiple leucine-rich repeats (LRRs) linked by a transmembrane domain to a cytosolic C-terminal intracellular signaling domain called Toll/IL-1R homology (TIR). TIR domain is named after its similarity to the intracellular domain of the interleukin-1 receptor (IL-1R) that mediates down-stream signaling events upon activation of the receptor (Jensen & Thomsen, 2012). TLR activation triggers a signal cascade involving adaptor proteins, protein kinases, and effector transcription factors, and mainly results in the production of type I IFN (e.g., IFN-beta). These processes induce an antiviral cellular state, providing an important first line of defense against virus infection. Others TLR-triggered proinflammatory cytokines (e.g. TNF and IL-6) are also important determinants of the balance between beneficial host innate immune responses and immunopathology.

Ten known TLR (TLRs 1–10) have been identified in humans. Among them TLR3, 7, 8 and 10 are known to recognize influenza virus infection (Koyama *et al.*, 2007; Le Goffic et al., 2007; Lee *et al.*, 2014). TLR7 and TLR8 share the highest degree of sequence similarity and sense ssRNA oligonucleotides containing guanosine and uridine-rich sequences. TLR7 is preferentially expressed in the endosomes of plasmacytoid dendritic cells (pDCs) and B cells, while TLR8 is preferentially expressed in myeloid DCs and monocytes. Upon stimulation, TLR7 and TLR8 recruit a TIR-containing adaptor named myeloid differentiation primary response gene 88 (MyD88) to the cytoplasmic TIR domain of the receptor (Medzhitov *et al.*, 1998). This signaling pathway leads to the activation of either NF-κB/AP-1 or NF-κB/AP-1/IRF7 complexes, which result in the transcription of proinflammatory cytokines and INF-α genes, respectively. TLR3 is localized in the intracellular compartment of macrophages, B lymphocytes, and DCs. It is located both intracellularly and on the surface of NK cells, epithelial cells, and fibroblasts. It recognizes viral dsRNA expressed in some viruses, leading to the activation of NF-κB/AP-1/IRF3 complexes and upregulation of IFN-β (Jensen & Thomsen, 2012). Recent studies reported an upregulation of human TLR10 expression upon influenza H5N1 viral infection, especially in primary macrophages. These data suggest that TLR10 is able to substantially enhance vRNP-induced activation of IL-8 expression (Lee *et al.,* 2014).

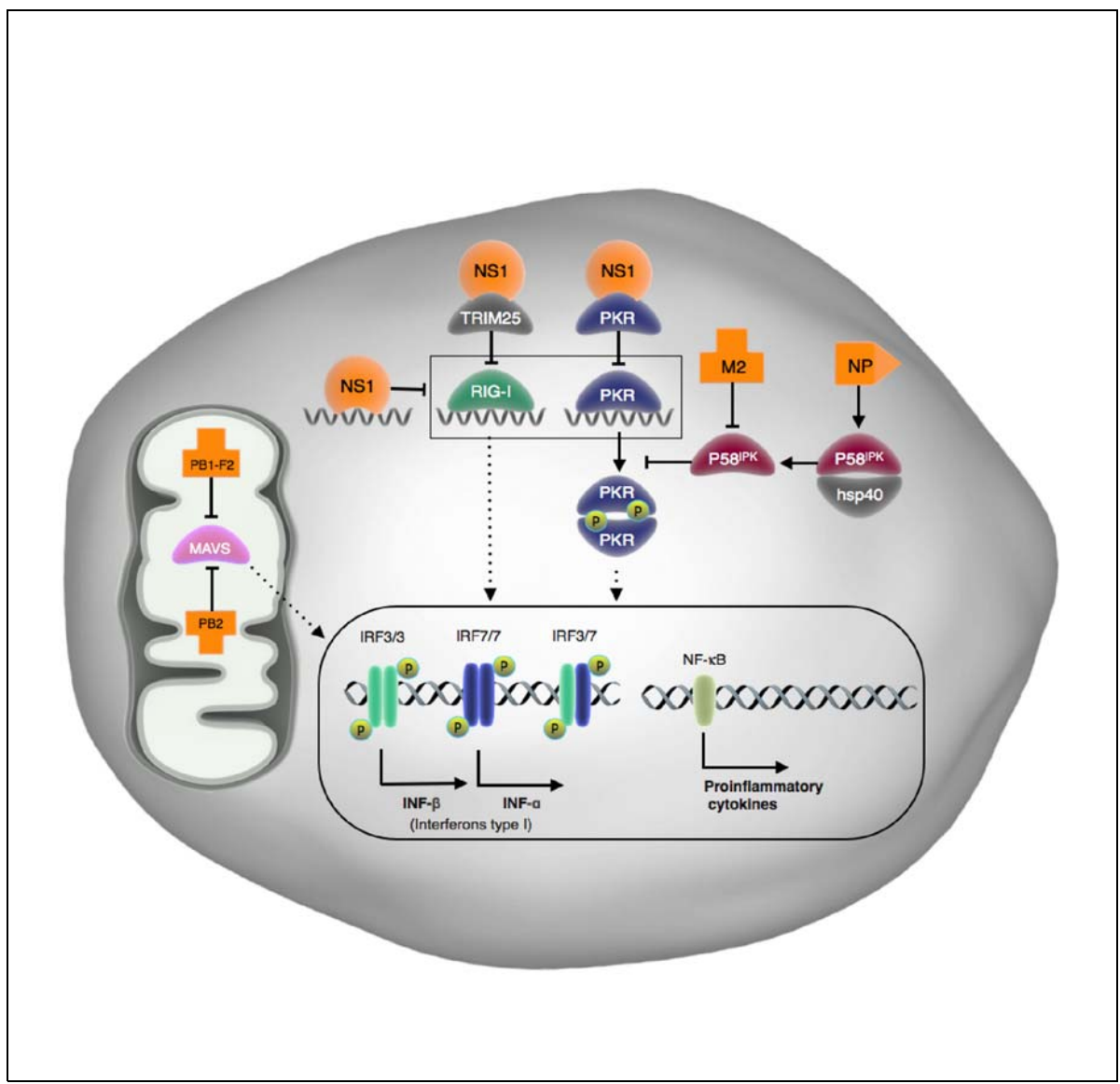

**Figure 2  Escape from adaptive response**.

Upon influenza virus infection, viral RNA is sensed and  several pathways are activated. This results in the production of type-I interferons, pro-inflammatory cytokines and induction of an antiviral state. Influenza virus proteins interact with this pathway. NS1 binds viral RNA to mask detection by PRRs. It blocks RIG-I activation by binding TRIM25 and also binds PKR and inhibits its phosphorylation, which occurs when PKR binds viral RNA. NP interacts with the P58IPK/hsp40 complex upon which P58ipk is dissociated from the complex and inhibits the phosphorylation of PKR. M2 on the other hand restores the P58IPK /hsp40 complex resulting in the inhibition of protein synthesis and induction of cell apoptosis, facilitating release of virus particles. PB1-F2 binds mitochondrial antiviral signaling protein (MAVS), thereby inhibiting this protein and type-1 interferon production. PB2 also binds MAVS and inhibits its function. Furthermore, PB2 binds interferon promoter stimulator 1 (IPS-1), resulting in inactivation of the promoter that would lead to production of type **I** interferon. Dotted arrows represent pathways and solid lines and arrows represent direct interactions.

### *Receptors II: RIG-I-like receptors (RLRs)*

The RIG-I-like receptors (RLRs) are cytosolic proteins that recognize viral RNA and are expressed by most human cell types **(Fig. 2)**. They share a conserved domain structure, characterized by a central DExD/H-box helicase domain and a ssRNA/dsRNA (ss/dsRNA)-binding C-terminal domain (CTD) which, when unbound, functions as a repressor domain (RD) (Gack, 2014). Three RLRs are known to be directly implicated in viral recognition: the retinoic acid-inducible gene I product (RIG-I, encoded by the DDX58 gene), the laboratory of genetics and physiology 2 (LGP2, encoded by the DHX58 gene) and the melanoma-differentiation-associated antigen 5 (MDA5). RIG-I and MDA5 harbor two N-terminal caspase activation and recruitment domains (CARDs) which, upon virus sensing, initiate downstream signaling, eventually leading to type I IFN gene expression. In contrast, LGP2 lacking the CARD signaling module has been shown to exert a regulatory role in RLR signaling, although its precise mechanism of action is still largely unknown (Kato *et al.*, 2006). RIG-I and LGP2 have been reported to play an important role in influenza virus RNA recognition, yet MDA5 recognizes also the viral RNA of other pathogens (Jensen & Thomsen, 2012). Recent studies revealed that RIG-I activity is tightly regulated by post-translational modifications; in particular, the dephosphorylation of CTD (Thr770, Ser854/855) and CARD (Ser8, Thr170) and their subsequent poly-ubiquitination at Lys63 and Lys172, respectively. Ubiquitin-bound CARDs facilitate RIG-I oligomerization and binding to MAVS, ultimately inducing antiviral signaling. (Gack, 2014).

### Receptors III: NOD-like receptors (NLRs)

Nucleotide-binding oligomerization domain–containing (NOD)-like receptors (NLR) are cytosolic proteins involved in the regulation of inflammatory and apoptotic responses, in particular during anti-viral responses (**Fig. 2**). NLRs contain a leucine-rich-repeat (LRR) domain located at the C-terminus and which is considered to be the RNA sensing region. The central NACHT domain mediates oligomerization and activation, while the N-terminal effector-binding domain, often a CARD or pyrin domain (PYD), upon activation and oligomerization of the whole NLR triggers the signal transduction cascade (Fritz *et al.*, 2006). Among NLRs, both NLRC2 (also named NOD2) and NLRP3 are involved in influenza virus RNA-sensing.

NLRC2, that contains two CARD domains at the N-terminus (Ogura *et al.*, 2001), is a NOD-like receptor family member that was recently shown to recognize ssRNA species derived from either one of RSV, parainfluenza virus and influenza-A virus. Upon viral ssRNA recognition, NLRC2 associates with MAVS (encoded by the VISA gene) through an interaction dependent on the LRR and nucleotide-binding domains (NBDs). This initiates the MAVS-dependent pathway (similar to the RLRs mechanism previously indicated) culminating with the expression of type-I IFN and proinflammatory cytokines. In contrast to RLRs, the NLRC2 interaction with MAVS doesn't involve CARD-CARD binding (Sabbah *et al.*, 2009). NLRP3 (for NOD-like receptor family pyrin domain-containing 3) contains a pyrin domain (PYD) that can interact with the N-terminal PYD domain of apoptosis-associated speck-like protein (ASC) containing a CARD domain. NLRP3 oligomerizes and recruits ASC and a procaspase-1 to form an inflammasome complex. This activates caspase-1, which subsequently mediates the conversion of pro-IL-1beta and pro-IL-18 to fully functional IL-1beta and IL-18. Activation occurs also upon infection with adenovirus, Sendai virus (ssRNA virus) and influenza virus A (Allen *et al.*, 2009; Ichinohe *et al.*, 2009).

### Interferon-induced proteins with tetratricopeptide repeats (IFITs)

IFITs are effectors within the innate immune system that seem to confer virus defense via disruption of protein–protein interactions in the host translation-initiation machinery. However, IFITs can directly recognize viral RNA bearing a 5'-triphosphate group (5'-PPP-RNA). IFIT1, and IFIT5 in particular, were shown to actively and selectively bind cytosolic 5'-PPP-ssRNAs. These results were confirmed by structural analysis (Pichlmair *et al.*, 2011; Abbas *et al.*, 2013).

### Viral escape from innate immunity

Two evolutionary processes are involved in viral evolution and in the evasion by the host immune system: antigenic drift of the influenza viral genome and the high selective pressure generated by the human immune response (**Fig. 2**). These escape mechanisms involve the interaction between protein viral effectors and some human targets. The influenza virus NS1 protein can bind viral RNA, masking it from TLRs/RIG-I recognition, thus inducing expression of INF type I (Garcia-Sastre, 2004; Guo *et al.*, 2007). NS1 proteins can also inactivate RIG-I binding to the tripartite-motif-containing protein 25 (TRIM25) and inhibit the ssRNA-binding protein kinase (PKR), encoded by the EIF2AK2 gene (Tan & Katze, 1998; Garcia-Sastre, 2004; Gack *et al.*, 2009). In addition, both the influenza virus nucleoprotein (NP) and the ion channel protein M2 can bind and inactivate the cellular inhibitor P58[IPK], encoded by the DNAJC3 gene, and the P58[IPK]/hsp40 complex respectively. These effectors are both critical for the regulation of PKR, and their inactivation results in the inhibition of protein synthesis, induction of cell apoptosis, and release of newly formed virus particles (Guan *et al.,* 2010).

**Table 2** presents the variability of host genes involved in viral escape mechanisms. The EFI2AK2 (another PKR) gene is the most variable gene among the four selected viral escape genes. As expected, the majority of variation is observed in the non-coding regions. In the coding region, the majority of the observed variation derives from single mutations (synonymous and non-synonymous).

**Table 2.** Host genes involved in viral escape mechanisms. **DNAJC3 (encodes the inhibitor P58[IPK]); EIF2AK2 (encodes PKR); TRIM25 (encodes the tripartite-motif-containing protein 25); VISA (encodes the MAVS protein)**

| Type of variant | DNAJC3 | EIF2AK2 | TRIM25 | VISA | SUM |
|---|---|---|---|---|---|
| Frameshift coding | 0 | 0 | 0 | 0 | 0 |
| Non-synonymous coding | 2 | 15 | 4 | 16 | 37 |
| Splice site | 0 | 8 | 0 | 4 | 12 |
| Synonymous coding | 5 | 16 | 3 | 6 | 30 |
| Intronic | 29 | 177 | 9 | 49 | 264 |
| 5' UTR | 2 | 24 | 1 | 0 | 27 |
| 3' UTR | 3 | 67 | 24 | 15 | 109 |
| Upstream | 2 | 5 | 0 | 0 | 7 |
| Downstream | 3 | 3 | 0 | 0 | 6 |
| Other* | 0 | 2 | 0 | 0 | 2 |
| **Total number of variants** | **46** | **315** | **41** | **90** | **492** |

The genetic variants described in the 1000 Genomes Project Pilot data are: *Frameshift coding:* structural mutation in coding sequence, resulting in a frame shift of reading (**A, B**); *Non-synonymous coding:* nucleotide substitution in the coding sequence (**A, B**), resulting in an amino acid change in the peptide chain (**A, C**); *Splice Site:* 1-3bp into an exon or 3-8bp into an intron (**A**); *Synonymous coding:* nucleotide substitution in coding sequence (**A, B**), but not resulting in an amino acid change (**C**); *Intronic:* mutation in intron (**A**); *5'UTR:* in 5' untranslated region (**A, B**); *3'UTR:* in 3' untranslated region (**A, B**); *Upstream:* mutation within 5kb upstream of the 5' end of transcript (**A**); *Downstream:* mutation within 5kb downstream of the 3' end of transcript (**A**).

*Mutations related to influenza infection*

Although the variability of the genes involved in innate immune response could have an important impact on the outcome of infection, our understanding of how genetic variability correlates with susceptibility to infection is still very limited and, in general, not supported by experimental approaches.

A missense mutation (F303S) in the Toll-like receptor 3 (TLR3) gene has been linked to influenza-associated encephalopathy (IAE), a severe neurological condition (Esposito *et al.*, 2012). F303S TLR3 receptor was shown to be less effective in activating the transcription factor, NF-κB, as well as triggering downstream signaling via the IFN beta receptor (Esposito *et al.*, 2012). An additional TLR3 SNP (rs5743313) was identified in a study of 51 children with confirmed H1N1 infection (Esposito *et al.*, 2012). The rs5743313C/T genotype was found in 18/18 children with IAV-associated pneumonia, but significantly less frequently in children with IAV without pneumonia (p < 0.0001), further demonstrating the association between TLR3 and IAV pathogenicity. Searching the 1000 Genomes Project database for such TLR3 variants, we found that none of the studied individuals carried the F303S mutation. In scope of this search we were not able to confirm that individuals with the

rs5743313C/T genotype went through acute pneumonia related to influenza-A infection (Esposito *et al.*, 2012). On the other hand, we were able to see that the frequencies of each allele were similar between populations, with the C allele having a frequency of more than 80%, suggesting that around 32% of individuals in the samples could possibly have the C/T genotype assuming that the population is in Hardy-Weinberg Equilibrium. Nonetheless, four populations of the 1000 Genomes Project were outside of this range of frequency for the C allele, namely—British (GBR, C=76% - 36% C/T genotype), Iberic (IBS, C=67% - 44% of C/T genotype), Puerto Ricans (PUR, 71% - 41% C/T genotype), and Toscani in Italy (TSI, C=69% - 43% C/T genotype), which indicates that those populations could be more susceptible to a more severe manifestation of the influenza infection, since the possibility of heterozygosity enhances when C and T allelic frequencies are similar. The Japanese population (JPT, C=97,8% - 4,3% C/T genotype) had the highest frequency of the C allele detected in the 1000 Genomes Project database, indicating that possibly only 4.3% of the population might be susceptible to acute pneumonia associated with IAV infection.

Four single nucleotide polymorphisms (SNPs) that showed association with severe illness were identified in a case-control association study on 91 patients with confirmed H1N1 infection–induced pneumonia (Mänz *et al.*, 2013). The first SNP was located in the genetic locus for the Replication Protein A Interacting Protein (RPAIN). RPAIN facilitates the nuclear localization of RPA, a key protein in maintaining DNA integrity and homeostasis. The second SNP was found in the Complement component 1 gene coding for a subcomponent binding protein (C1QBP) gene. C1QBP inhibits the complement activation. The third SNP was found within the gene encoding the CD32 (also called FCGR2A). The CD32 is a cell surface receptor with low affinity for IgG, it is found on phagocytic cells such as macrophages and neutrophiles that are involved in the process of phagocytosis and clearing of immune complexes. The last SNP was located within the potentially intergenic region of chromosome 3. Since two of these SNPs are associated with the genes whose products take part in either the clearing of immune complexes (FCGR2A), or in complement activation (C1QBP), it is reasonable to hypothesize that the severe disease outcome of H1N1 infection may result from variation in the host's immune response.

A new study reports a direct association between polymorphisms of rs17561 in the IL1A gene and rs1143627 in the IL1B gene and higher susceptibility to A(H1N1) infection (Liu *et al.*, 2013). Interleukins IL1A and IL1B are very important in influenza infection, as both induce the expression of a variety of inflammatory mediators that may initiate the cascade of inflammatory responses and activation of T cells (Acosta-Rodriguez *et al.*, 2007). IL1B has been demonstrated to mediate acute pulmonary inflammation, which is one of the main causes of death related to influenza-A virus, while IL1A secretion is regulated by the NLRP3 inflammasome (Schmitz *et al.*, 2005; Dawood *et al.*, 2012; Gross *et al.*, 2012). The rs17561T variant in the IL1A gene is present in low frequencies in the 1000 Genomes Project samples. The frequency of this allele ranges from 6% (in Southern Han Chinese) to a maximum of 35% (in British samples). The European samples (Iberian, British, Europeans residents in Utah, Toscani and Finnish) showed the highest frequencies of the T allele associated with the higher susceptibility to H1N1 influenza-A infection. The SNP rs1143627C in the IL1B gene is very frequent in all populations in the 1000 Genomes Project, but in particular among Mexicans, Puerto Ricans, Luthya from Kenya, and Yoruba from Nigeria, ranging from 52% in Mexicans to 75% in Luhya. In other populations, the frequencies range from 21% (in Iberians), to 48% (in Americans of African descent). If such SNPs strongly influence susceptibility to influenza-A infection, the high frequencies might explain rapid spread and the acute respiratory inflammation among human populations.

The mechanisms of RNA sensing and pattern recognition are very important in the first phase of infection. Mutations in the genes that encode pattern-recognizing proteins could prevent a successful recognition of pathogen patterns. It is striking that a low number of mutations in these genes is observed, as shown in **Table 3**. Most of the variation is located in intronic regions, which may be neutral to the function of the expressed products. This is not true for the TLR genes, where both non-synonymous and synonymous modifications are predominantly in coding regions, which may probably reflect evolutionary increase in the types of patterns recognized.

**Table 3**. List of variants within genes involved in the innate immune response to influenza virus infection (the data retrieved from the 1000 Genomes Project database)

| Type of variant | DDX58 | DHX58 | NLRP3 | NOD2 | IFIT1 | IFIT5 | TLR3 | TLR7 | TLR8 | TLR10 | SUM |
|---|---|---|---|---|---|---|---|---|---|---|---|
| Frameshift coding | 3 | 1 | 0 | 2 | 1 | 0 | 1 | 1 | 0 | 2 | 11 |
| Non-synonymous coding | 30 | 8 | 42 | 28 | 5 | 5 | 10 | 11 | 10 | 34 | 183 |
| Splice site | 3 | 3 | 3 | 0 | 0 | 1 | 1 | 0 | 2 | 0 | 13 |
| Synonymous coding | 24 | 1 | 54 | 13 | 6 | 2 | 10 | 6 | 33 | 22 | 171 |
| Intronic | 124 | 22 | 234 | 16 | 0 | 2 | 10 | 1 | 0 | 12 | 421 |
| 5' UTR | 12 | 2 | 40 | 3 | 1 | 0 | 4 | 0 | 1 | 8 | 71 |
| 3' UTR | 43 | 0 | 55 | 10 | 1 | 26 | 1 | 13 | 27 | 12 | 188 |
| Upstream | 1 | 0 | 3 | 1 | 6 | 1 | 1 | 0 | 4 | 6 | 23 |
| Downstream | 5 | 0 | 5 | 0 | 3 | 1 | 1 | 0 | 2 | 8 | 25 |
| Other* | 1 | 0 | 0 | 0 | 0 | 0 | 1 | 1 | 0 | 4 | 7 |
| Total number of variants | 246 | 37 | 436 | 73 | 23 | 38 | 40 | 33 | 79 | 108 | 1113 |

*Essential Splice Site (DDX58); Stop Gained (TLR3, TLR7, TLR10). The genetic variants described in the 1000 Genomes Project Pilot data are: *Frameshift coding:* structural mutation in coding sequence, resulting in a frame shift of reading (A, B); *Non-synonymous coding:* nucleotide substitution in the coding sequence (A, B), resulting in an amino acid change in the peptide chain (A, C); *Splice Site:* 1-3bp into an exon or 3-8bp into an intron (A); *Synonymous coding:* nucleotide substitution in coding sequence (A, B), but not resulting in an amino acid change (C); *Intronic:* mutation in intron (A); *5'UTR:* in 5' untranslated region (A, B); *3'UTR:* in 3' untranslated region (A, B); *Upstream:* mutation within 5kb upstream of the 5' end of transcript (A); *Downstream:* mutation within 5kb downstream of the 3' end of transcript (A).

## Infection spread: adaptive immune response to influenza virus

The adaptive phase of the immune response to influenza virus is crucial for the effective fight against infection, and can be divided into humoral and cellular immune response (**Fig. 3**). The humoral immune response is characterized by the degradation of viral proteins into peptides, further shown on the cell surface of antigen-presenting cells (APC) in complex with human leucocyte antigen (HLA) class-II proteins. The HLA class-II–epitope complexes interact with CD4+ T-cells, which in turn trigger the production of virus-specific antibodies. In contrast to the humoral response, the cellular immune response results in the activation of naïve CD8+ T cells and their differentiation into cytotoxic T lymphocytes (CTLs). Immunoproteasomes present in the host cell cytosol degrade viral proteins into peptides that are then transported to the endoplasmic reticulum (ER) via TAP (transporter of antigen processing), where they are loaded onto HLA class-I molecules. HLA class-I–epitope complexes are then transported via the Golgi to the cell membrane, where they can be recognized by virus-specific CTLs (Guermonprez *et al.*, 2002). The CTLs are able to kill the infected cells, thus limiting the spread of the virus.

### Humoral immune response to influenza virus

Elderly people exposed to influenza-A/H1N1 virus in the 1950s were relatively spared from infection during the 2009 pandemic strain, since they have developed cross-reactive antibodies (Yu *et al.*, 2008; Hancock *et al.*, 2009). This observation suggests that the humoral immune system triggers long-lasting antibody-mediated protection against influenza strains that resemble the original strain of infection. In the humoral adaptive phase of the immune response, B-cells are stimulated by the production of influenza-virus-specific antibodies primarily directed against HA, NA and M2 viral proteins (Baumgarth *et al.*, 2000; Waffarn & Baumgarth, 2011). Antibodies directed against the trimeric globular HA head efficiently block receptor-mediated endocytosis of the virus (De Jong *et al.*, 2003). However, most of antibodies are strain-specific and fail to neutralize other influenza serotypes (Yu *et al.*, 2008).

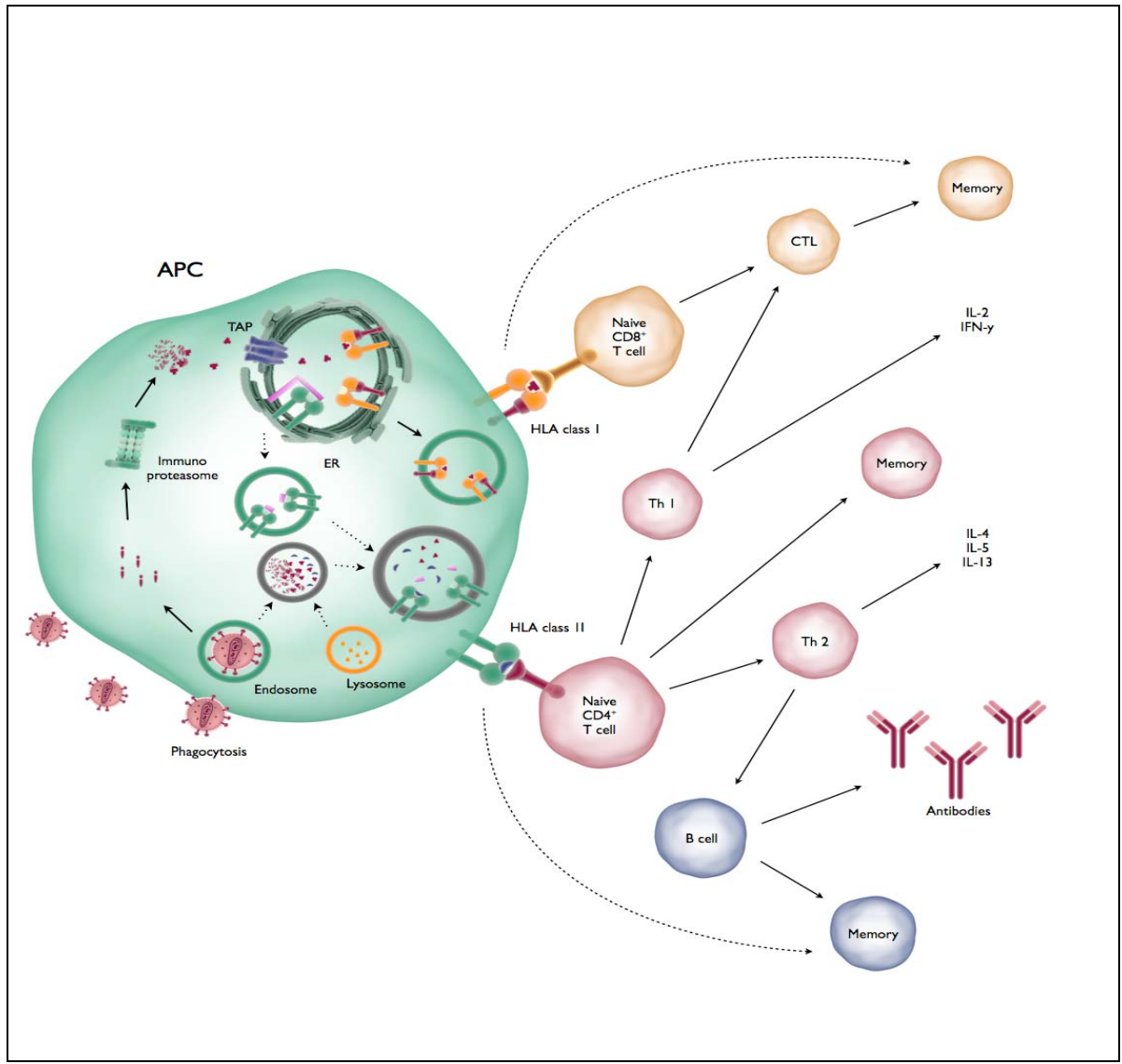

**Figure 3   HLA antigen presentation pathways and activation of the adaptive immune response**

Both HLA class I and HLA class II presentation pathways are activated during influenza virus infections. HLA class I presentation (solid arrows): the epitopes obtained by immunoproteasome degradation of cytosol proteins, are transported in the ER via TAP. The HLA class I epitope complexes are then transported to the plasma membrane where they stimulate the differentiation of cytotoxic CD8+ T cell (CTL) and provide signaling for the suppression of the infected cells by CTL effector (adaptive cellular response). HLA class II presentation (dotted arrows): the epitopes obtained from the proteolytic degradation within endosomal compartments are then assembled with HLA class II proteins. The HLA class II - epitope complexes are transported to the plasma membrane where they both stimulate the differentiation of T-helper CD4+ and activate T-helper effectors triggering the humoral response. Induction of immune responses after a primary influenza-A virus infection is indicated by solid arrows. Virus-specific memory cell populations upon secondary encounter with an influenza-A virus are indicated by dashed arrows.

Neuraminidase (NA) is a viral glycoside hydrolase which, catalyzing the hydrolysis of terminal sialic acid residues from both newly formed virions and host cell receptors, facilitates the release and spread of the newly formed viral particles  (von Itzstein, 2007). Anti-NA antibodies inhibit the enzymatic activity of NA thereby limiting viral spread and shorten severity and duration of illness (Bosch *et al.*, 2010; Kilbourne *et al.*, 2009). NA-specific antibodies may also contribute to the clearance of virus-infected cells by facilitating Antibody-Dependent Cell-Mediated Cytotoxicity (ADCC) (Carolien *et al.*, 2012). Tetrameric transmembrane viral protein M2 has ion channel activity and plays an important role in unpacking the virus in the endosome (Schnell & Chou, 2008). The protective effect of cross-reactive anti-M2 antibodies was first demonstrated in mice after passive transfer of M2-specific monoclonal antibodies (Kim *et al.*, 2013), and is attributed to ADCC activation (Mozdzanowska *et al.*, 1999; El Bakkouri *et al.*, 2011). The heterosubtypic protective effect of M2-specific antibodies most probably stems from strong conservation of M2 protein sequence among different influenza virus strains (Tompkins *et al.*, 2007; Fiers *et al.*, 2009; Fu *et al.*, 2009; Schotsaert *et al.*, 2009; Wang *et al.*, 2009; Ebrahini & Tebianian, 2011).

The humoral response induces the production of antibodies targeting other viral proteins, including the highly conserved NP protein (Carragher *et al.*, 2008). Although NP-specific antibodies are non-neutralizing to viruses, there is some evidence that they contribute to establishing protective immunity (Lamere *et al.*, 2011; Carragher *et al.*, 2008). Their mode of action may include ADCC of infected cells and opsonisation of NP, resulting in improved T-cell responses (Bodewes *et al.*, 2010; Sambhara *et al.*, 2001). However, the protective effect of non HA/NA antibodies is still under debate.

The IgA, IgG and IgM antibody isotypes are induced upon primary influenza virus infection, whereas IgM responses are not observed after secondary infection (Carolien *et al.*, 2012). Virus-specific serum IgA responses seem indicative for recent infection with influenza virus (Voeten *et al.*, 1998; Koutsonanos *et al.*, 2011), while virus-specific IgG antibodies correlate with long-lasting protection, whereby these antibodies match the strains causing the infection (Koutsonanos *et al.*, 2011; Onodera *et al.*, 2012). IgM antibodies can neutralize the virus but also activate the complement system (Jayasekera *et al.*, 2007; Fernandez Gonzalez *et al.*, 2008). In addition to serum antibodies, influenza virus infection also induces local mucosal secretory IgA antibody responses aimed at protecting airway epithelial cells from infection (Onodera *et al.*, 2012).

### *Cellular immune response to influenza virus – T Cells CD4+.*

Activation of CD4+ T cells occurs upon recognition of viral epitopes associated with MHC class-II molecules and interaction with co-stimulatory molecules on APCs. Naïve CD4+ T cells can differentiate into CD4+ Th1 cells or Th2 cells (**Fig. 3**). Th1 cells are mainly involved in regulation of the CTL response (Zhu & Paul, 2010a), and in the induction of memory CD8+ T cells (Riberdy *et al.*, 2000; Belz *et al.*, 2002; Deliyannis *et al.*, 2002). Th2 cells promote the activation and differentiation of B-cells, resulting in antibody production (Zhu & Paul, 2010b; Okoye & Wilson, 2011). Two different processes, namely antibody class-switching and somatic hypermutation determine the affinity maturation of the influenza-specific antibodies, drastically empowering their efficiency (Kamperschroer *et al.* 2006). Memory CD4+ T cells, induced after a primary influenza-A virus infection, contribute to a faster control of subsequent influenza-A virus infections (Strutt *et al.*, 2010). Lung resident memory CD4+ T cells in particular have an important role in protection against secondary influenza-A infection.

### *Cellular immune response to influenza virus – T Cell CD8+.*

The activation of CD8+ T cells occurs upon recognition of viral epitopes associated with MHC class-I molecules on APCs in the draining lymph nodes. This process leads to the differentiation of CD8+ T into cytotoxic T lymphocytes (CTLs), which migrate to the site of infection in order to recognize and eliminate the host cells infected by the influenza virus, thereby preventing the production and spread of progeny virus (Nakanishi *et al.*, 2009). The interaction of T-cell receptor (TCR) with the HLA class-I–epitope complex stimulates CTLs to release perforin and granzymes, causing the apoptosis of infected cells (Andrade, 2010; Moffat *et al.*, 2009). Proinflammatory cytokines like TNF-alpha, which inhibit virus replication and enhance lytic activity (La Gruta *et al.*, 2004; La Gruta *et al.*, 2007; Metkar *et al.*, 2008; Andrade, 2010; van Domselaar & Bovenschen, 2011), are also produced. After infection, the formation of a pool of memory CD8+ T cells occurs, constituting the basis for strong and fast recall responses upon secondary infections (Zimmermann *et al.*, 1999; Chang *et al.*, 2007; Hikono *et al.*, 2007; DiSpirito & Shen, 2010; van Gisbergen *et al.*, 2011). Virus-specific CTLs are mainly directed against epitopes present within highly conserved viral proteins, such as NP, M1 and PA. This implies an efficient cross-reactivity that allows the CTLs to attack cells infected by different influenza-A subtypes (Assarssen *et al.*, 2008; Kreijtz *et al.*, 2008; Lee *et al.*, 2008). Studies on mice confirmed this hypothesis, showing that CD8+ T cells contribute to both homo- and heterosubtypic immunity (Kreijtz *et al.*, 2007; Grebe *et al.*, 2008; Kreijtz *et al.*, 2009; Hillaire *et al.*, 2011a; Hillaire *et al.*, 2011b).

However, the evidence of CTL-induced immune protection against influenza in human is still very limited. The presence of heterosubtypic memory CD4+ and CD8+ T cells against the 2009 pandemic H1N1 virus was detected in naïve individuals (Sridhar *et al.*, 2012). More circumstantial evidence for a protective role for CD8+ T cells in heterosubtypic influenza infections stems from epidemiological studies. People who had a symptomatic influenza-A infection with the H1N1 strain prior to the 1957 pandemic were partially protected from infection with the pandemic H2N2 strain. A similar trend was found in isolated infections with the H5N1 (Carolien *et al.*, 2012).

### Genetic variability in the adaptive immune response

The genetic variability of HLA genetic loci is known to be very large in any given population (*c.f.* **Table 3**). HLA hyper-variability is in fact the basic mechanism used by the host to ensure the recognition of potentially dangerous non-self molecules, triggering activation of the adaptive immune response against these molecules (**Tables 4A and 4B**). The recognition of the HLA-epitope complexes by T cell receptor (TCR) is crucial for activation of the adaptive immune response. With this degree of genetic variability, it is extremely difficult to associate single HLA mutations with a differential response to influenza virus infection. Usually entire HLA alleles or specific allele combinations are considered, as they are more likely to be associated with specific phenotypic effects. Mutations in these regions could either prevent the formation of HLA-epitope complexes (Price *et al.*, 2000; Voeten *et al.*, 2000; Berkhoff *et al.*, 2004; Rimmelzwaan *et al.*, 2004; Berkhoff *et al.*, 2007a) or cause the epitope to escape recognition by TCR, since the epitope no longer matches the specificity of the TCR (Price *et al.*, 2000; Boon *et al.*, 2002; Berkhoff *et al.*, 2007a; Berkhoff *et al.*, 2007b).

**Table 4A.** Genetic variability within the three principal genetic *loci* of classic HLA class-I genes (the data retrieved from the 1000 Genomes Project database)

| Type of variant | HLA-A | HLA-B | HLA-C | SUM |
|---|---|---|---|---|
| Essential splice site | 34 | 10 | 12 | **56** |
| Stop Gained | 71 | 26 | 23 | **120** |
| Frameshift coding | 577 | 127 | 94 | **798** |
| Non-synonymous coding | 1232 | 382 | 857 | **2471** |
| Splice site | 206 | 28 | 48 | **282** |
| Synonymous coding | 486 | 148 | 339 | **973** |
| Intronic | 1519 | 557 | 1230 | **3306** |
| 5' UTR | 74 | 28 | 8 | **110** |
| 3' UTR | 909 | 191 | 268 | **1368** |
| Upstream | 60 | 14 | 34 | **108** |
| Downstream | 70 | 60 | 59 | **189** |
| Total number of variants | 5133 | 1535 | 2937 | **9605** |

The genetic variants described in the 1000 Genomes Project Pilot data are: *Frameshift coding:* structural mutation in coding sequence, resulting in a frame shift of reading (**A, B**); *Non-synonymous coding:* nucleotide substitution in the coding sequence (**A, B**), resulting in an amino acid change in the peptide chain (**A, C**); *Splice Site:* 1-3bp into an exon or 3-8bp into an intron (**A**); *Synonymous coding:* nucleotide substitution in coding sequence (**A, B**), but not resulting in an amino acid change (**C**); *Intronic:* mutation in intron (**A**); *5'UTR:* in 5' untranslated region (**A**); *3'UTR:* in 3' untranslated region (**A, B**); *Upstream:* mutation within 5kb upstream of the 5' end of transcript (**A**); *Downstream:* mutation within 5kb downstream of the 3' end of transcript (**A**).

Mutations within viral epitopes were shown to have a direct effect on the efficiency of CTL response (Carolien *et al.*, 2012). These types of mutations have been observed in escape mutants of viruses that chronically infect their host, like HIV-1 (Huet *et al.*, 1990; Cale *et al.*, 2011). Substitutions of epitopes recognized by CTL have also been observed during the evolution of seasonal A/H3N2 influenza viruses (Voeten *et al.*, 2000; Rimmelzwaan *et al.*, 2004; Berkhoff *et al.*, 2007a). The R384G substitution in the HLA-B*2705-restricted NP383–391 epitope (Voeten *et al.*, 2000) considerably affects the human virus-specific CTL response *in vitro* (Berkhoff *et al.*, 2007a) while the HLA-B*3501 restricted NP418–426 epitope (Boon *et al.*, 2002; Berkhoff *et al.*, 2007b) displays signs of antigenic drift (Boon *et al.*, 2004), explaining the cross-reactivity of CTL against contemporary viruses with historic strains (Gras *et al.*, 2010). The variation rate of CTL epitopes could be considerably functionally constrained. For instance, the M1 58–66 epitope is highly conserved despite its immunodominant nature, probably because mutations within this epitope compromise the virus fitness (Berkhoff *et al.*, 2006).

**Table 4B.** Genetic variability of HLA class II β-chains within the three principal genetic *loci* (the data retrieved from the 1000 Genomes Project database)

| Type of variant | HLA-DPB1 | HLA-DQB1 | HLA-DQB2 | HLA-DRB1 | HLA-DRB3 | HLA-DRB5 | SUM |
|---|---|---|---|---|---|---|---|
| Essential splice site | 2 | 27 | 0 | 2 | 3 | 5 | **39** |
| Stop Gained | 12 | 42 | 0 | 4 | 1 | 11 | **70** |
| Frameshift coding | 0 | 150 | 0 | 35 | 3 | 24 | **212** |
| Non-synonymous coding | 176 | 1071 | 18 | 166 | 89 | 150 | **1670** |
| Splice site | 10 | 138 | 4 | 9 | 2 | 2 | **165** |
| Synonymous coding | 69 | 491 | 33 | 90 | 38 | 62 | **783** |
| Intronic | 192 | 3099 | 181 | 267 | 142 | 191 | **4072** |
| 5' UTR | 3 | 60 | 1 | 16 | 13 | 4 | **97** |
| 3' UTR | 117 | 633 | 48 | 113 | 86 | 130 | **1127** |
| Upstream | 7 | 100 | 3 | 20 | 15 | 11 | **156** |
| Downstream | 9 | 141 | 4 | 28 | 17 | 6 | **205** |
| Total number of variants | 597 | 5883 | 292 | 744 | 405 | 580 | **8501** |

The genetic variants described in the 1000 Genomes Project Pilot data are: *Frameshift coding:* structural mutation in coding sequence, resulting in a frame shift of reading (**A, B**); *Non-synonymous coding:* nucleotide substitution in the coding sequence (**A, B**), resulting in an amino acid change in the peptide chain (**A, C**); *Splice Site:* 1-3bp into an exon or 3-8bp into an intron (**A**); *Synonymous coding:* nucleotide substitution in coding sequence (**A, B**), but not resulting in an amino acid change (**C**); *Intronic:* mutation in intron (**A**); *5'UTR:* in 5' untranslated region (**A, B**); *3'UTR:* in 3' untranslated region (**A, B**); *Upstream:* mutation within 5kb upstream of the 5' end of transcript (**A**); *Downstream:* mutation within 5kb downstream of the 3' end of transcript (**A**).

It is noteworthy that the most variable of the non-HLA genes (NLRP3, with 496 variants) are still less variable, according to the data from the 1000-genomes Project, than the average for the nine HLA genes shown here. The most conserved gene in the HLA system, HLA-DQB2, has 292 described variants. The HLA-B gene, on the other hand, has 1,535 variants, which is still lower than that recorded by the IPD-TGTM/HLA website, which as (of May 2014) identifies 3,455 alleles, encoding 2,577 different proteins. Such enormous quantity of variants, underscores the importance of their function, since they should attach and present a plethora of possible peptides to cytotoxic and memory lymphocytes. The more universal are the HLA molecules, the more efficient will be the peptide presentation and the activity of the adaptive immune system.

## Concluding remarks

**It is more important to know what sort of person has a disease, than to know what sort of disease a person has.** – *Hippocrates of Cos (c. 460 BC – c. 370 BC)*

The initial motivation for this review was the work of Horby *et al.* (2012), that was concluded with an open question: *Is susceptibility to severe influenza in humans heritable? – hard to answer, but not due to a lack of genotyping or analytic tools, nor because of insufficient evidence from severe* *influenza cases, but because of the absence of a coordinated effort to define and assemble cohorts of cases*. Here we presented an update overview of the host variability of genes associated with flu infection, and discuss the possible influence of this variability on the severity of flu infection. Each virus-host interaction begins with entry of the virus into the host organism, and ends either in death of the individual or in elimination of the virus by the immune system. Evolution of the host–pathogen interactions yields variants in host genes, several of which are ultimately associated with bad or good prognosis following infection, while boosting the host's future resilience towards given, specific strain subtype of the viral epitopes. Even if the severity of infection was mild and/or asymptomatic, these variants have been shown to be polymorphic in different human populations, which could be correlated with the different rates of morbidity/mortality from the influenza-A infectivity. Therefore, the answer to the Horby *et al.* (2012) question is positive, the susceptibility to severe influenza in humans is heritable.

On the other hand, it does not mean that all causes and variance in symptomatology and morbidity of flu infections have been discovered, or that they are only due to host genetic profile, but there are clearly several genes in the human genome that might have a direct impact on the course of influenza-A infections. In particular, when considering the ethnicity-derived genomic variance, some aspects of specific habitats and climates should not be neglected. In temperate regions the peak flu infectivity is felt predominantly during the winter months and epidemics recur with a highly predictable seasonal pattern. The studies using mouse-adapted strains of influenza virus, with experiments performed in the winter months yielded a transmission rate of 58.2%, yet of only 34.1% in summer (Schulman and Kilbourne 1963). It has been hypothesized that possible causes of such seasonality include fluctuations in host immune efficiency mediated by seasonal factors such as melatonin (Dowell 2001) and vitamin D (Cannell *et al.* 2006) levels; the changes in host behavior (school attendance, indoor crowding, air travel, etc.); but in particular environmental factors, including temperature (Eccles 2002), relative humidity, and the direction of seasonal winds movement in the upper atmosphere (Hammond *et al.* 1989). The role of the South-East Asia region, in particular in countries spanning latitudes from subtropical at the south to high-mountainous at the north through several climactic zones, as spawning grounds of new flu strains is very well known (see e.g. Le *et al.* 2013), and a field of intense phylogeographical research (Wallace and Fitch 2008, for a review). In contrast, the situation differs in the [sub]tropical regions, which experience influenza throughout the year, with somewhat increased incidence during rainy seasons (Viboud *et al.* 2006, Shek and Lee 2003). Such regions, like rainforests of equatorial Africa, or Amazonian regions of Brazil, Venezuela and Ecuador, are still potentially rich areas for a concerted efforts to connect the possible ethnic genomics factors in isolated communities, with the influenza-A prevalence and infectivity patterns.

Also, many factors can cause modifications in DNA (epigenetic modifications) that alter gene expression. Several studies support the theory of prenatal/childhood programming as origin of various adult diseases (Nicoletto and Rinaldi, 2011, Ahmed 2010). Epigenetic modifications were described to explain adult disease in response to environmental modifications in prenatal or childhood. For example, Yang and Huffman (2013) showed that individuals who suffer some sort of nutritional privation during gestation tend to become obese adults. It can be thought that if the pattern of methylation that is programmed in one's genome during development can affect the way that individual will react to a diet as an adult, it might possibly affect the way immune-related genes are expressed during infections (Guoying W *et al.* 2013).

Very recently the work of Guihot *et al.* (2014) demonstrated some curious cases that impose more questions than answers. The authors have found biological markers that can predict whether an individual is more prone to suffer lethal consequences to influenza infection. However, the genetic variability that underlies such markers is not well understood – e.g., some of the markers indicated a "trapping" antibodies in the lungs for the lethal cases, whilst the survivors had the antibodies freed into a blood. In another recent study Worobey *et al.* (2014) have hypothesized that the high mortality during the 1918 Spanish flu pandemic among adults aged ~20 to ~40 years might have been due primarily to their childhood exposure to a heterosubtypic putative H3N8 virus – estimated as to have been circulating from ~1889–1900. All other age groups (except immunologically naive infants) were likely to be partially protected by childhood exposure to N1 and/or H1-related antigens. Such tenet poses immediate questions. What possible immunological mechanism[s] already described, might explain such an increase of mortality rates? Especially a mechanism of contrasting the H3 non-naïve young-adult Spanish flu victims (but putatively naïve towards H1 serotype), with infants – by definition naïve to all HA subtypes. In effect such a mechanism would require a convincingly plausible pathway[s] of suppressing immunity specifically by an action of H3-derived epitopes. One possible mechanism would be that the exposure to this putative H3N8 virus could have caused a methylation pattern in some genes, especially on some paths of imune response, and as consequence these patients had an extremely severe flu. Also, maybe not all individuals exposed to this virus developed the same methylation patterns and final supressions due to the genetic variability known to exist in human populations. The authors postulate that a mechanism akin to original antigenic sin (OAS, Francis T, 1960) may have interfered with immune responses in some of those infected in 1918, and peaked in those exposed previously to the 1889 virus. However, such hypothesis opens important questions about possible molecular and immunological mechanisms pertaining to the actual OAS phenomena "in action". It is obvious that many more studies of this type will be necessary to better understand how the environment and host genetic variability affects the susceptibility to the influenza-A virus. The present review shows that both host genetic variability as well as environment are clearly factors of great importance to determine such susceptibility.

## Acknowledgements

We are grateful to Anna Lisa Lucido (Jackson Laboratory for Genomic Medicine) for her critical review of the manuscript. This work was supported by the Polish Ministry of Education and Science (grant number NCN 2013/09/B/NZ2/00121). GM was financed by research fellowship within Project ''Information technologies: Research and their interdisciplinary applications'',. ACA was supported by a doctoral fellowship granted by Coordenação de Aperfeiçoamento Pessoal de Nível Superior, CAPES., Brazil. SFO received a fellowship from CNPq (Conselho Nacional de Desenvolvimento Científico), Brazil. Additional partial funding was generously provided by the WND-POIG.01.01.02-00-007/08 grant from the European Regional Development Fund.

## Conflicts of interests

Authors declare no conflict of interests.